\documentclass[sigconf,screen]{acmart}

\setcopyright{cc}
\setcctype{by}
\acmDOI{10.1145/3832783.3837426}
\acmYear{2026}
\copyrightyear{2026}
\acmISBN{979-8-4007-2882-2/2026/10}
\acmConference[ASE '26]{Proceedings of the 41st IEEE/ACM International Conference on Automated Software Engineering}{October 12--16, 2026}{Munich, Germany}
\acmBooktitle{Proceedings of the 41st IEEE/ACM International Conference on Automated Software Engineering (ASE '26), October 12--16, 2026, Munich, Germany}
\acmSubmissionID{ase26main-p208-p}
\received{2026-03-26}
\received[accepted]{2026-06-18}

\usepackage{graphicx}
\usepackage{multirow}
\usepackage{makecell}
\usepackage{listings}
\usepackage{fancyvrb} 
\usepackage{fontawesome} 
\usepackage{enumitem}
\usepackage{pifont}
\usepackage{graphicx}   
\usepackage{soul}         
\usepackage{xspace}
\usepackage{xcolor}
\usepackage{balance}

\newcommand{\dx}[1]{\textcolor{black}{{#1}}}
\newcommand{\appname}{{RACE-Bench}\xspace}

\begin{document}

\title{RACE-Bench: A Reasoning-Augmented Benchmark for Repository-Level Code Agents on Feature Addition}

\author{Shuhan Liu}
\authornote{Both authors contributed equally to this research.}
\orcid{0009-0000-2848-3735}
\affiliation{
  \institution{Zhejiang University}
  \city{Hangzhou}
  \country{China}
}
\email{liushuhan@zju.edu.cn}

\author{Zhiyi Zhao}
\authornotemark[1]
\orcid{0009-0002-3564-6605}
\affiliation{
  \institution{Zhejiang University}
  \city{Hangzhou}
  \country{China}
}
\email{zhaozhiyi@zju.edu.cn}

\author{Xing Hu}
\authornote{Corresponding authors: Xing Hu and Kui Liu}
\orcid{0000-0003-0093-3292}
\affiliation{
  \institution{Zhejiang University}
  \city{Hangzhou}
  \country{China}
}
\email{xinghu@zju.edu.cn}

\author{Kui Liu}
\authornotemark[2]
\orcid{0000-0003-0145-615X}
\affiliation{
  \institution{Independent researcher}
  \city{Shanghai}
  \country{China}
}
\email{brucekuiliu@gmail.com}

\author{Xiaohu Yang}
\orcid{0000-0003-4111-4189}
\affiliation{
  \institution{Zhejiang University}
  \city{Hangzhou}
  \country{China}
}
\email{yangxh@zju.edu.cn}

\author{Xin Xia}
\orcid{0000-0002-6302-3256}
\affiliation{
  \institution{Zhejiang University}
  \city{Hangzhou}
  \country{China}
}
\affiliation{
  \institution{Hangzhou High-Tech Zone (Binjiang) Institute of Blockchain and Data Security}
  \city{Hangzhou}
  \country{China}
}
\email{xin.xia@acm.org}

\begin{abstract}
Repository-level code agents have shown strong promise in real-world feature addition tasks, making reliable evaluation of their capabilities increasingly important. However, existing benchmarks primarily evaluate these agents as black boxes based on final test correctness, providing limited insight into how they reason and where failures arise. To address this limitation, we introduce \appname, a reasoning-augmented benchmark for evaluating code agents on repository-level feature addition tasks. \appname contains 528 real-world feature addition instances from 12 open-source repositories. Each instance is paired with executable patch verification and structured intermediate \dx{reference reasoning} covering issue understanding, file localization, implementation tasks, and step decomposition. Based on this design, we introduce a dual-track evaluation framework that jointly measures patch correctness and intermediate reasoning \dx{alignment with developer-accepted reference trajectories}.
We evaluate three representative repository-level code agents on \appname. On the full benchmark, \texttt{Resolved Rate} ranges from 29\% to 70\% across different agents. Our reasoning-level analysis further shows that while current agents perform well at understanding high-level intent, their performance degrades substantially when translating intent into concrete implementation steps. We also find 
\dx{patches that can be applied but still fail the tests cover fewer reference-reasoning elements (35.7\% lower recall) and contain more unsupported reasoning elements (94.1\% higher over-prediction) than successful patches.}
These findings highlight the importance of evaluating repository-level code agents beyond final patch correctness by examining the quality of their reasoning processes.
\end{abstract}



\maketitle

\vspace{-0.1cm}

\section{Introduction}
\label{sec:introduction}
Recent years have witnessed the rapid growth of large language models (LLMs), giving rise to powerful models such as ChatGPT~\cite{achiam2023gpt} and Gemini~\cite{team2023gemini}. These models have quickly reshaped automated software engineering and are now widely applied to code generation and related development tasks~\cite{guo2024deepseek,roziere2023code,lin2024llm,yang2024ecosystem,liu2025creme}. This technological shift has influenced real-world development practices: a substantial proportion of developers regularly rely on LLM-based tools (e.g., GitHub Copilot~\cite{githubcopilot}) to write, modify, or review code~\cite{wermelinger2023using,barke2023grounded,dakhel2023copilot}. These tools significantly improve productivity and reduce cognitive load during programming activities~\cite{joshi2023repair,vaithilingam2022expectation}.

Beyond single-turn code generation, recent research has introduced code agents that extend LLMs into autonomous, multi-step problem-solving systems capable of interacting with code repositories~\cite{gao2025trae,zhang2024autocoderover,wang2024openhands}. Unlike LLMs that generate code in isolation, code agents iteratively reason about a given problem, invoke external tools (e.g., code search, testing frameworks, and execution environments), and adapt their behavior based on intermediate feedback. They can perform a broad range of repository-level development tasks, including fixing bugs and implementing feature requests~\cite{bouzenia2024repairagent,zhang2024autocoderover,wang2024openhands,yang2024swe}, generating tests from natural-language issue descriptions~\cite{mundler2024code,ahmed2024tdd,nashid2025issue2test}, and automating complex project setups~\cite{hu2025repo2run,bouzenia2025you}.
\dx{Among these tasks, feature addition is a distinct setting for code agents. Unlike bug fixing, which often starts from an observed failure and asks the agent to restore intended behavior, feature addition requires the agent to infer new externally visible behavior, decide how to integrate it into existing APIs or module boundaries, and preserve existing behavior while adding the new capability.} This trend highlights the need to systematically evaluate the capabilities of code agents in feature addition scenarios.

However, the failure modes of code agents are more complex compared with traditional code-generation LLMs. In real-world feature addition workflows, producing a correct patch is the outcome of a structured reasoning pipeline that involves \ding{182} understanding the issue and its requirements, \ding{183} identifying the relevant files, \ding{184} locating the necessary code entities, and \ding{185} formulating an appropriate modification plan. Errors can arise at any stage of this process: an agent may misunderstand the problem intent, modify irrelevant files, introduce unnecessary changes, or omit essential steps. Moreover, code agents may occasionally pass test cases despite relying on coincidental reasoning, leading to apparent success that hides underlying problems. As a result, evaluating code agents solely based on whether the final patch passes tests, without examining the intermediate reasoning process, offers limited insight into their true capabilities and failure characteristics.

Existing benchmarks for evaluating code LLMs and code agents fall short in capturing the intermediate reasoning challenges in repository-level feature addition scenarios. Function-level benchmarks such as HumanEval~\cite{chen2021evaluating}, MBPP~\cite{austin2021program}, and LiveCodeBench~\cite{jain2024livecodebench} focus on small, context-independent function generation tasks. While effective for evaluating isolated code generation capabilities, these benchmarks fail to reflect real-world development, where changes often span across multiple interdependent files and code entities. To bridge this gap, recent benchmarks such as SWE-bench~\cite{jimenez2023swe} and FEA-bench~\cite{li2025fea} derive tasks from real issues and patches in open-source repositories. However, although these benchmarks represent a step toward more realistic evaluation, they focus solely on whether the final generated patch passes the tests. This black-box evaluation paradigm offers limited visibility into an agent’s intermediate reasoning process and makes it difficult to diagnose where and why an agent fails within the development workflow.

To address this gap, we present a new benchmark named \textbf{\appname}
(Reasoning-Augmented Code Agent Evaluation) for systematically evaluating 
code agents in realistic repository-level feature addition scenarios with explicit reasoning evaluation.
\appname is designed with the following key characteristics:
\ding{192} \appname is constructed from high-quality, actively maintained open-source repositories, with each instance grounded in a real feature request and its corresponding pull request (PR).
\ding{193} \appname focuses on feature addition tasks, which require agents to infer intended functionality from natural-language issues and coordinate changes across multiple files and code entities.
\ding{194} \appname provides structured \dx{developer-accepted reference trajectories} for intermediate reasoning stages, including issue understanding, relevant file identification, concrete implementation tasks, and abstract task-step decomposition.
\ding{195} Each task is equipped with the \emph{Fail-to-Pass (FTP)} and \emph{Pass-to-Pass (PTP)} test cases, verifying both the correctness of generated code and the preservation of original functionality.
\ding{196} \appname enables fine-grained evaluation beyond final patch correctness by measuring reasoning recall and over-prediction (e.g., redundant reasoning steps) at different modules of the development workflow.
\ding{197} \appname provides a unified, containerized execution environment for each instance using Docker, enabling reliable reproduction of repository states and consistent evaluation across different agents.
Overall, \appname contains 528 feature addition instances derived from 12 actively maintained open-source GitHub repositories. 
To facilitate more efficient experimentation, we construct a lightweight subset named \appname Lite. This subset contains 100 instances selected through a difficulty-aware sampling strategy designed to preserve the overall challenge level of the \appname.

To evaluate the effectiveness of \appname, we assess three representative code agents on the full benchmark using \texttt{DeepSeek} as the base model. The results reveal substantial performance variation across agents, with \texttt{Resolved Rate} ranging from 29\% to 70\%.
\dx{Beyond final patch correctness, our reasoning evaluation shows that code agents usually understand high-level intent, but their reference-alignment recall drops as reasoning becomes more concrete, from file localization to task steps. Furthermore, patches that can be applied but still fail the tests show 35.7\% lower reference-reasoning recall and 94.1\% higher over-prediction than successful patches.}
We further extend the evaluation on \appname Lite by incorporating additional base LLMs. The results indicate that although the choice of base LLM influences agent performance, its impact is constrained by the agent architecture. Agents with better-designed frameworks maintain relatively stable performance across different base LLMs.

\noindent\textbf{Contributions:} In summary, the main contributions of this paper can be summarized as follows:
\vspace{-0.05cm}
\begin{itemize}[leftmargin=*]
    \item We introduce \appname, a repository-level feature-addition benchmark for code agents. It contains 528 real-world feature addition instances from 12 GitHub repositories, each equipped with executable patch verification and structured intermediate \dx{\emph{Reference Reasoning}}.
    \item We propose a dual-track evaluation framework that measures both patch correctness and \dx{reasoning alignment with developer-accepted trajectories} through five structured reasoning modules, enabling fine-grained diagnosis beyond patch-only evaluation.
    \item  We conduct a systematic study of three representative code agents with multiple base LLMs on \appname, revealing differences in effectiveness, reasoning behavior, and efficiency.
\end{itemize}

\section{\appname}
\label{sec:racebench}
\begin{figure*}[h]
  \centering  \includegraphics[width=\linewidth]{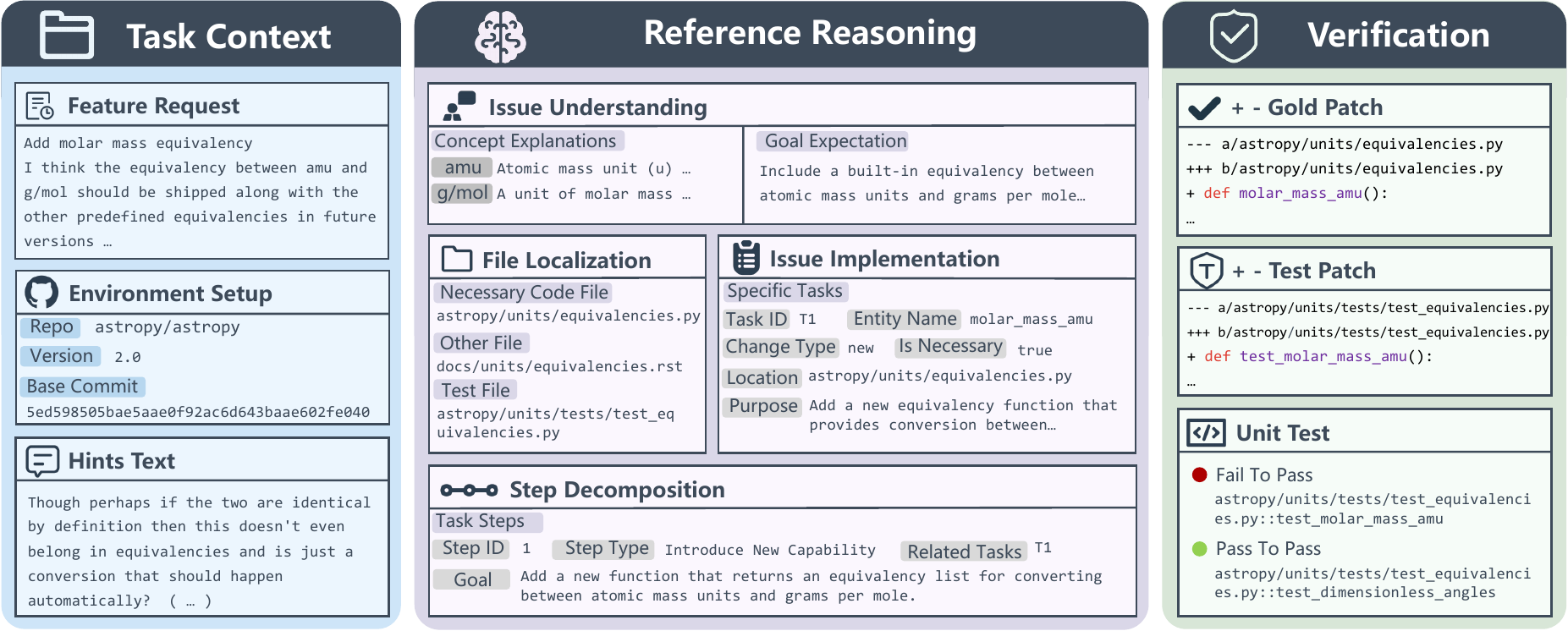}
  \vspace{-1.5em}
  \caption{An example of a task instance from the \appname. During agent inference, only the \emph{Feature Request} and \emph{FTP} tests are provided as the observable inputs, while the \emph{Environment Setup} specifies the execution context for evaluation. Each instance additionally includes structured \dx{developer-accepted \emph{Reference Reasoning}}, the corresponding \emph{Gold Patch}, and \emph{Unit Tests}, enabling end-to-end evaluation of both intermediate reasoning processes and final patch correctness.}
\label{fig:instance}
\vspace{-1.5em}
\end{figure*}

In this section, we present the design and construction of \appname.

\subsection{Overview}
\label{sec:overview}
The task instances in \appname are constructed from real-world PR data and their corresponding GitHub issues. As illustrated in Figure~\ref{fig:instance}, an instance supports the evaluation of both final patch correctness and \dx{reasoning alignment}. Each instance consists of three primary components:

\noindent\ding{182} \textbf{Task Context. }This module provides the necessary information and execution environment to initialize a development task. The \emph{Feature Request} is derived from the original GitHub issue and provides a natural language description of the intended functionality. The \emph{Environment Setup} specifies the execution context, including the repository, release version, and base commit. This setup ensures that evaluations are conducted in a consistent and reproducible state. Additionally, the module may include optional \emph{Hints Text} from developer discussions to provide extra context for design decisions.

\noindent\ding{183} \dx{\textbf{Reference Reasoning.}} 
To evaluate the \dx{reasoning alignment} of code agents, \appname follows a human-like workflow that decomposes the reasoning process into four stages containing five reasoning modules:
\begin{itemize}[leftmargin=*] \item \textbf{Issue Understanding.} This component captures the understanding of the issue intent. The \emph{Concept Explanations} identify domain-specific concepts, while the \emph{Goal Expectation} provides a concise summary of the expected behavior.

\item \textbf{File Localization.} This component identifies the files that are modified or added. All files are categorized into three types. \emph{Necessary Code Files} are essential for the patch to pass the tests. \emph{Other Files} include auxiliary changes like documentation or changelogs. \emph{Test Files} consist of the unit test files for the task instance.

\item \textbf{Issue Implementation.} 
This component identifies the code entities to be added, modified, or removed. For each entity, it includes the file location and the change type (i.e., \texttt{new}, \texttt{modified}, or \texttt{deleted}), along with its \emph{purpose} and an \emph{is\_necessary} flag to indicate whether it is an indispensable entity to pass the tests. \dx{This level diagnoses whether an agent covers the concrete code entities required by the developer-accepted implementation.}

\item \textbf{Step Decomposition.} This component abstracts the implementation process into a sequence of essential task steps. Each step describes a high-level modification action (e.g., \texttt{introduce new capability} and \texttt{reuse existing semantics}).
\dx{It evaluates whether an agent produces a reference-aligned implementation plan.}
\end{itemize}

\dx{The five modules follow a human-like workflow for feature addition. This granularity balances diagnostic value and reliability. A coarser schema would hide failure sources, while a finer schema would depend more on implementation details and reduce annotation consistency.}
\dx{The reference reasoning in \appname is derived from developer-accepted patches and represents one accepted implementation trajectory rather than the only valid solution.}

\noindent\ding{184} \textbf{Verification. }This module evaluates the functional correctness of a generated patch through automated testing. Each instance provides a \emph{Gold Patch} and an associated \emph{Test Patch} for the requested feature from the PR. These patches can be applied to the repository using the \texttt{git apply} tool. The correctness is verified by executing the \emph{Unit Test} with \texttt{pytest}. The \emph{FTP} test cases fail before applying the \emph{Gold Patch} and pass afterward. The \emph{PTP} test cases ensure that existing functionalities remain preserved after the modification.

\subsection{Benchmark Construction}
\label{sec:benchmarkConstruction}
This subsection describes the end-to-end construction pipeline of \appname, including repository selection and environment setup, instance filtering, \dx{\emph{Reference Reasoning}} construction, and the validation process to ensure data quality and reliability.

\subsubsection{Repository Collection and Environment Construction}
\label{sec:experimentdocker}
Following the design of SWE-bench~\cite{jimenez2023swe}, we select the same set of 12 popular open-source Python repositories on GitHub as the source for \appname instances. These repositories are widely used in practice, actively maintained, and governed by well-defined contribution guidelines.
To ensure that both instance construction and subsequent benchmark evaluation are performed in a stable and reproducible environment, we build independent \emph{Environment Docker} images for each stable release of the repositories. Each image encapsulates the required dependencies and Python runtime. This approach mitigates environment drift and ensures that tests are executed consistently across experiments.

\subsubsection{Instance Filtering}
Starting from all merged PRs across the selected 12 repositories, we apply a three-stage filtering pipeline to identify high-quality feature addition instances suitable for evaluating repository-level code agents.

\noindent \textbf{Attribute-based filtering.}
We first perform coarse-grained filtering based on observable PR attributes. Specifically, we retain only PRs that (1) modify files under test-related directories (e.g., \texttt{test} or \texttt{testing}), indicating the presence of explicit test changes, and (2) are explicitly linked to at least one GitHub issue, ensuring the availability of a natural-language feature request. After this stage, 14,333 PRs satisfy the filtering criteria.

\noindent \textbf{Intent-based filtering.}
Next, we identify PRs that correspond to feature addition tasks. For each remaining PR, we use \texttt{DeepSeek} to classify the development intent based on the PR title and description, assigning it to categories such as \emph{new feature}, \emph{bug fix}, or \emph{refactor}. 
To assess the reliability of the classifier, we randomly sample 100 PRs from the filtered pool and manually verify their predicted labels. The manual inspection shows that the classification accuracy exceeds 99\%, indicating that the LLM provides reliable intent predictions for this task.
We retain only PRs whose intent is classified as \emph{new feature} and manually recheck the candidate PRs to confirm that they correspond to feature addition tasks. This filtering step yields 2,931 candidate PRs.

\noindent \textbf{Execution-based filtering.}
Finally, we apply an execution-driven validation process to ensure that each candidate PR can serve as a reliable benchmark instance. For each PR, we retrieve its corresponding base commit and repository version, and perform the following steps within the matching \emph{Environment Docker}:
(1) clone and install the repository;
(2) check out the base commit;
(3) apply the \emph{Test Patch} and execute the test suite, where at least one test is expected to fail;
(4) apply the \emph{Gold Patch} and re-run the tests;
(5) compare the test outcomes before and after applying the \emph{Gold Patch} to determine the \emph{FTP} and \emph{PTP} test sets.
If any step fails (e.g., environment setup errors or missing \emph{FTP} tests), the instance is discarded. After this stage, 580 PRs remain and are retained as valid \appname instances.

\subsubsection{\dx{Reference Reasoning} Construction}
\label{sec:reasoninggroundtruthconstruction}
For each retained instance, we construct a structured \dx{developer-accepted \emph{Reference Reasoning}} to enable evaluation of intermediate reasoning behaviors exhibited by code agents.
The \dx{reference} is organized as the four-stage reasoning schema described in Section~\ref{sec:overview}, and is derived through a combination of LLM assistance and patch-driven program analysis. \dx{All LLM calls in this construction stage use \texttt{DeepSeek} with temperature=0.0. The prompts are included in the replication package.}

\noindent \textbf{Issue Understanding.}
The \emph{Concept Explanations} and \emph{Goal Expectation} are generated using LLM. We provide the model with the original \emph{Feature Request}, the \emph{Hints Text}, and the repository \texttt{README} to supply task-specific and project-level context. The model identifies domain-specific concepts from the issue description and summarizes the expected system behavior after implementation. These summaries focus on observable behavior. To prevent implementation leakage, we explicitly prohibit code-level details.

\noindent \textbf{File Localization.}
The \emph{File Localization} \dx{reference} is extracted from the \emph{Gold Patch} and \emph{Test Patch}.
All files modified or added in the \emph{Gold Patch} are initially marked as relevant. To distinguish necessary files from others, we employ a test-driven ablation procedure. For each modified file, we create a patch variant by removing the changes of that file and then re-executing the \emph{Unit Test}. A file is labeled as a \emph{Necessary Code File} if removing its changes causes a test failure; otherwise, it is labeled as an \emph{Other File}. In addition, all files modified or added in the \emph{Test Patch} are labeled as \emph{Test File}.

\noindent \textbf{Issue Implementation.}
The \emph{Specific Tasks} \dx{reference} captures the set of code entities required to implement the feature. For each instance, we analyze the \emph{Gold Patch} by statically parsing the modified source files to extract added, modified, or deleted code entities (i.e., functions and methods). We define classes as a structural container rather than a code entity, as they typically consist of multiple methods. For each extracted entity, we further annotate its semantic \emph{purpose} using \texttt{DeepSeek}. Specifically, we provide the model with the original \emph{Feature Request} and the patch-local code context of the entity, and instruct the LLM to summarize the role of the entity in terms of observable behavior. To determine whether an entity is semantically necessary, we additionally provide the \emph{FTP} tests and ask the model to assess whether the intended behavior can still be achieved without modifying that entity.

\noindent \textbf{Step Decomposition.}
Finally, we construct the \emph{Task Steps} \dx{reference} by abstracting the gold
implementation into a minimal sequence of essential semantic steps.
We define a fixed and closed taxonomy of step types
(i.e., \texttt{introduce new capability}, \texttt{reuse existing semantics},
\texttt{change existing semantics}, \texttt{deprecate or replace behavior} and \texttt{enforce constraints or edge cases}).
This taxonomy captures common categories of externally observable implementation changes required for tests to pass.
Given the filtered \emph{Specific Tasks}, the \emph{Gold Patch}, and the original \emph{Feature Request}, \texttt{DeepSeek} is prompted to select and order a minimal set of task steps from this taxonomy. Each step represents exactly one irreducible semantic responsibility, is grounded in one or more existing tasks, and is assigned a single step type. This representation abstracts low-level code details while preserving the essential structure and ordering of the implementation process.

\subsubsection{Validation}
\label{sec:human_validation}
To ensure the reliability of the automatically constructed \dx{\emph{Reference Reasoning}}, we perform an \emph{LLM-assisted audit} over 580 candidate instances that remain after instance filtering, followed by targeted human inspection on a subset of cases. \dx{The audit is conducted using \texttt{GPT-5.2} with temperature 0.0, and the full audit prompt is provided in the replication package.}

For each candidate instance, a \texttt{GPT 5.2} LLM auditor evaluates three aspects: 
(1) For \emph{Issue Understanding}, we assess whether the \emph{Concept Explanations} accurately cover the domain-specific terms appearing in the \emph{Feature Request} and whether the \emph{Goal Expectation} correctly summarizes the intended requirement.
(2) For \emph{Specific Tasks}, we verify whether entities labeled as necessary correspond to genuine requirements of the \emph{Feature Request} and whether their inferred purposes are semantically accurate.
(3) For \emph{Step Decomposition}, we assess whether each \emph{Task Step} represents a distinct and essential semantic step by performing an ablation check. In this process, a step is considered valid only if its removal renders the \emph{Feature Request} unsatisfiable.

Instances are removed if the auditor detects semantic misinterpretation, incorrect necessity judgments, or structural defects. 
\dx{To assess the reliability of the automatic audit, we manually inspect 100 candidates, including 50 accepted and 50 rejected by the auditor. Two authors independently review the \emph{Feature Request}, patches, \emph{Reference Reasoning}, and the auditor rationale. They accept a case if the labels are semantically supported. They reject cases with unsupported additions, incorrect necessity labels, or invalid step decompositions. For example, \texttt{pydata\_\_xarray-9453} is accepted because its reference reasoning covers all required behaviors. In contrast, \texttt{sympy\_\_sympy-12171} is rejected because its feature request includes both Derivative printing and Float exponent conversion, but the reference omits the Float requirement.}
On this subset, the human inspectors agree with the auditor’s decision on 96 out of 100 cases (96.0\%), suggesting that the automatic audit is generally reliable for filtering problematic \dx{\emph{Reference Reasoning}}. In total, 52 out of the 580 candidate instances are excluded. After this stage, 528 validated instances remain and form the full \appname.

\begin{table*}[t]
\centering
\small
\caption{Reasoning evaluation metrics for different intermediate reasoning modules.}
\vspace{-1em}
\begin{tabular}{p{2.8cm} p{3.5cm} p{10cm}}
\toprule
\textbf{Reasoning Module} & \textbf{Metric} & \textbf{Definition} \\
\midrule
\multirow{3}{*}{\textbf{Concept Explanations}}
& Recall@Concept 
& Proportion of benchmark concepts that are explicitly explained by the agent. \\
& \multirow{2}{*}{Accuracy@Concept}
& Proportion of agent-explained concepts whose explanations are judged semantically consistent with the benchmark by an LLM judge. \\
\midrule
\multirow{2}{*}{\textbf{Goal Expectation}}
& \multirow{2}{*}{Score@GoalExpectation}
& A 10-point LLM-based evaluation of whether the agent correctly understood the core user intent. \\
\midrule
\multirow{2}{*}{\textbf{Relevant Files} }
& Recall@RelevantFiles 
& Coverage of necessary files in the benchmark that are correctly identified by the agent. \\
& OverPrediction@RelevantFiles 
& Proportion of irrelevant files predicted by the agent to all predicted files. \\
\midrule
\multirow{3}{*}{\textbf{Specific Tasks} }
& Recall@SpecificTasks 
& Proportion of necessary implementation tasks covered by the agent. \\
& \multirow{2}{*}{OverPrediction@SpecificTasks}
& Proportion of extra tasks predicted by the agent that are not required by the gold patch to all predicted tasks. \\
\midrule
\multirow{2}{*}{\textbf{Task Steps} }
& Recall@TaskSteps 
& Proportion of benchmark steps that are recovered by the agent. \\
& OverPrediction@TaskSteps 
& Proportion of extra steps predicted by the agent to all predicted steps. \\
\bottomrule
\end{tabular}
\label{tab:reasoning_metrics}
\vspace{-0.5em}
\end{table*}

\subsubsection{\appname Lite Construction}
\label{sec:liteset}

Evaluating repository-level code agents requires substantial LLM interaction, making large-scale evaluation costly. To enable efficient experimentation while preserving the overall difficulty distribution of the benchmark, we construct a lightweight subset, \appname Lite.

Starting from the validated instances in the full benchmark, we select a subset of size $N$ (default $N{=}100$) whose empirical difficulty closely matches that of the full benchmark. Specifically, for each instance we collect two results from the full benchmark runs of three agent configurations (AutoCodeRover, TraeAgent, and mini-SWE-Agent): (1) whether the generated patch passes the test suite (\emph{patch success}) and (2) whether the patch can be applied successfully (\emph{apply success}). Let $r^{patch}_k$ and $r^{apply}_k$ denote the \texttt{Resolved Rate} and \texttt{Patch Apply Rate} of configuration $k$ on the full benchmark.
Given a candidate subset $S$, we compute the corresponding rates $\hat r^{patch}_k(S)$ and $\hat r^{apply}_k(S)$. The subset selection objective minimizes the deviation from the full benchmark statistics:

\[
\min_{S} \sum_{k}
\Big(
(\hat r^{patch}_k(S) - r^{patch}_k)^2 +
(\hat r^{apply}_k(S) - r^{apply}_k)^2
\Big).
\]

We solve this optimization using a randomized local-search procedure. This process iteratively replaces instances in the subset to reduce the objective value.
The resulting \appname Lite preserves the \texttt{Resolved Rate} and \texttt{Patch Apply Rate} distributions observed in the full benchmark while reducing evaluation cost. We use \appname Lite for larger-scale experiments across multiple models in RQ3 (Section~\ref{rq:rq3}).

\begin{figure}[h]
  \centering  \includegraphics[width=\linewidth]{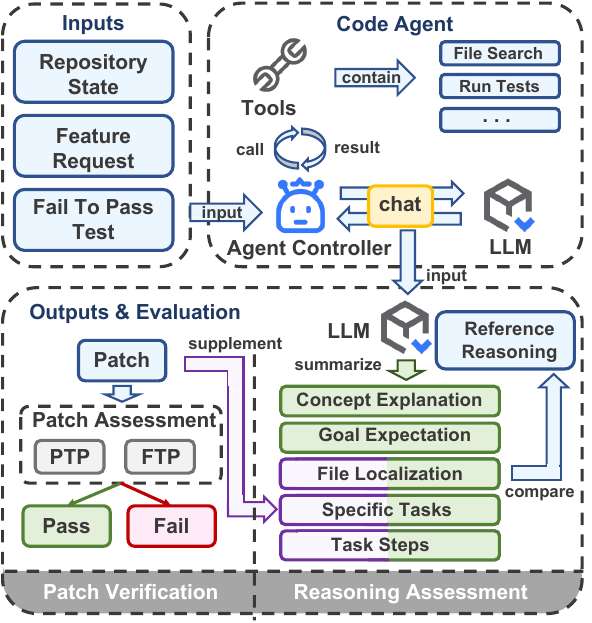}
  \vspace{-1.5em}
  \caption{The Dual-Track Evaluation Framework for \appname.}
\label{fig:workflow}
\vspace{-1em}
\end{figure}

\subsection{Evaluation Workflow}
\label{sec:evaluationWorkflow}
We propose a dual-track evaluation workflow to systematically assess both the final patch output and the intermediate reasoning process of code agents. As illustrated in Figure~\ref{fig:workflow}, the workflow consists of three stages: agent inputs, code agent execution, and dual-track evaluation.

\noindent\textbf{Agent Inputs.}
A code agent is initialized in the repository state defined by the \emph{Environment Setup} in Figure~\ref{fig:instance}. \dx{\appname targets feature addition tasks with developer-provided tests. The agent receives the \emph{Feature Request} and all \emph{FTP} tests as inputs, while \emph{PTP} tests remain hidden for final evaluation. The \emph{FTP} tests serve as an executable specification of the requested behavior, defining the minimal correctness criteria that the generated patch must satisfy.}

\noindent \textbf{Code Agent Execution.}
A code agent typically operates through the coordination of an agent controller, external development tools, and an underlying LLM.
Commonly used tools include file search, static analysis, and test execution.
Guided by the LLM’s intermediate outputs, the agent iteratively invokes these tools to inspect the codebase, validate hypotheses, and gather task-relevant information.
This reasoning–action loop continues until the agent generates a final patch.
The interaction traces between the agent controller and the LLM record the decisions and rationales underlying the code changes, and serve as the primary evidence for analyzing the agent’s reasoning behavior.

\noindent \textbf{Dual-Track Evaluation.}
The evaluation of an instance consists of two complementary components:
\begin{itemize}[leftmargin=*]
\item \textbf{Patch Verification:}
The agent is expected to generate a patch that satisfies all \emph{FTP} tests while preserving existing behaviors validated by \emph{PTP} tests.
A patch is considered functionally correct if it passes all required tests in the evaluation environment.

\item \textbf{Reasoning Assessment:} 
We employ a summarizer LLM to post-process the interaction traces between the agent controller and the underlying LLM.
These traces are converted into a structured representation that aligns with the predefined \dx{\emph{Reference Reasoning}} schema.
If the agent successfully produces a valid patch, we further enrich the summarized reasoning by applying the same patch-driven back-inference procedure used to construct the \dx{\emph{Reference Reasoning}}, extracting additional information such as file localization, specific implementation tasks, and abstract task steps.
Finally, the resulting structured reasoning is systematically compared with the \dx{\emph{Reference Reasoning}}.
Detailed evaluation metrics are presented in Section~\ref{sec:evaluationMetrics}.
\end{itemize}

\subsection{Evaluation Metrics}
\label{sec:evaluationMetrics}
We evaluate agents on the proposed benchmark from two complementary perspectives: \textbf{\emph{Patch Evaluation}} and \textbf{\emph{Reasoning Evaluation}}. \emph{Patch Evaluation} focuses on the correctness of the generated code, while \emph{Reasoning Evaluation} assesses the quality of the intermediate reasoning produced during the problem-solving process.

\noindent \textbf{\emph{Patch Evaluation}.} 
An agent is considered successful on an instance if the generated patch passes all \emph{FTP} and \emph{PTP} tests. Following prior work on repository-level code generation benchmarks~\cite{li2025fea,chen2025featbench,jimenez2023swe,deng2025nocode}, we adopt \textbf{Resolved Rate (\%)} as the primary metric for evaluating patch correctness. This metric measures the percentage of instances for which the agent produces a successful patch on the first attempt.
We also report the \textbf{Patch Apply Rate (\%)}, which captures the proportion of generated patches that are syntactically valid and can be applied to the repository without errors.
\dx{Apply-success is checked before test execution. Patches that fail to apply with \texttt{git apply} are classified as apply-fail and are not evaluated on the FTP/PTP tests.}

\noindent \textbf{\emph{Reasoning Evaluation}.} Table~\ref{tab:reasoning_metrics} presents the evaluation metrics used for each reasoning module. \dx{Since feature addition may admit multiple valid implementations, these metrics should be interpreted as alignment with the developer-accepted reference reasoning.} 
For a subset of the intermediate reasoning modules that require semantic judgment, we employ an LLM-based evaluation paradigm (\emph{LLM-as-Judge}). The reliability of \emph{LLM-as-Judge} has been widely studied and shown to be effective not only in natural language generation tasks~\cite{liu2023g,song2024finesure}, but also in software engineering-related tasks~\cite{he2025code,wang2025can}. 
\dx{All LLM-based reasoning evaluation steps use \texttt{DeepSeek} with temperature 0.0. The prompt and scripts are included in the replication package.}
To compute the metrics for \emph{Specific Tasks}, it is necessary to align benchmark tasks with the tasks inferred from the agent-generated patch. We adopt a hybrid alignment strategy that combines deterministic matching with semantic matching.
For tasks whose \emph{Change Type} is \texttt{deleted} or \texttt{modified}, we apply deterministic alignment: a task is considered aligned if both its \emph{Entity Name} and \emph{Location} match exactly between the benchmark and the agent output.
For tasks with \emph{Change Type} \texttt{new}, we first attempt deterministic matching using the same criteria. If no aligned task is found, we further employ an LLM-based semantic matcher to compare the \emph{Purpose} descriptions and identify an aligned task at the functional level.
A similar alignment procedure is applied to the \emph{Task Steps} module. For each benchmark step, we first select candidate agent steps that share the same \emph{Step Type} and have overlapping \emph{Related Tasks}. We then use an LLM-based semantic judgment to determine whether any candidate step covers the benchmark step’s intent.
\dx{To validate the evaluation procedure, two authors manually reviewed 100 evaluated agent runs. They marked an evaluation as correct when the reported match or score was supported by both the \emph{Reference Reasoning} and the agent trace. They marked it as incorrect when the evaluator missed a semantically equivalent match, accepted an unrelated match, or assigned a score that was not supported by the evidence. Disagreements were resolved through discussion. The authors agreed with the automatic evaluation in 97 of the 100 runs, indicating that the procedure is reliable for measuring alignment with the \emph{Reference Reasoning}.}

\section{Experimental Setup}
\label{sec:experimentSetUp}
\subsection{Agents and Models}
\noindent \textbf{Agent selection.}
We evaluate three representative repository-level code agents selected from the SWE-bench leaderboard~\cite{swebenchleaderboard2024}, which provides a widely used reference point for current automated software engineering agents.
Specifically, we include AutoCodeRover~\cite{zhang2024autocoderover}, a research system that emphasizes structured repository navigation and localized edits; TraeAgent~\cite{gao2025trae}, which investigates test-time scaling and iterative refinement for software engineering tasks; and mini-SWE-Agent~\cite{yang2024swe}, a lightweight agent framework derived from SWE-Agent that integrates LLM reasoning with repository execution environments.
We select these agents because they are widely used in SWE-bench evaluations, represent both research-oriented and practical agent designs, and have publicly documented workflows that enable reproducible evaluation.

\noindent \textbf{Model selection.}
We use \texttt{DeepSeek V3.2}~\cite{liu2024deepseek} as the base model and evaluate it across all three agents on the full \appname.
To broaden model coverage under a lower-cost setting, we additionally report the results of three agents on \appname Lite using \texttt{GLM-4.7}~\cite{glm47} and \texttt{GPT-5.2}~\cite{openai_gpt5} as base models.
In addition, \texttt{DeepSeek V3.2} is used as the summarizer LLM to generate a normalized intermediate reasoning chain for the reasoning evaluation.

\subsection{Setup}
\noindent \textbf{Execution environment.}
To ensure reproducibility, we evaluate each instance in an isolated \emph{Instance Docker} built on the \emph{Environment Docker} image of the related release version described in Section~\ref{sec:experimentdocker}.
For every instance, we reset the repository to its base commit, clone the target project, and perform the required project setup steps (e.g., installing the package with \texttt{pip install -e .}).

\noindent \textbf{Hardware.}
All experiments are conducted on a Linux server equipped with two Intel Xeon Platinum 8358P CPUs (64 cores total @ 2.60~GHz) and 2~TB RAM, running Ubuntu 20.04 LTS.

\noindent \textbf{Agent configuration.}
For all agents, we use their default configurations, following the recommendations in the official documentation.
To compute \texttt{Resolved Rate}, we run each agent on each instance exactly once (i.e., single attempt per instance).

\noindent \textbf{LLM configuration.}
All LLM calls in this work are made through official APIs.
We set \texttt{temperature}=0 and \texttt{top\_p}=1.
The maximum generation length is set to 4,096 tokens for agent inference and 8,192 tokens when the model is used as the summarizer LLM for intermediate reasoning normalization.

\section{Results and Analysis}
\label{sec:results}
\label{subsec:results_overview}
We organize our evaluation around three research questions (RQs):

\noindent\textbf{RQ1. (Capability)} How well do code agents resolve repository-level feature addition tasks on \appname?

\noindent\textbf{RQ2. (Reasoning)} How well do code agents perform on intermediate reasoning evaluation, and what does the reasoning profile reveal about their strengths and limitations?

\noindent\textbf{RQ3. (Generality)} How does the choice of the agent's base LLM affect performance on \appname Lite?

\subsection{RQ1: Capability}
\label{rq:rq1}
RQ1 evaluates the patch-level capability of code agents to resolve repository-level feature addition tasks.
We evaluate three agents on the full \appname and measure their end-to-end effectiveness by whether the generated patch can (i) be applied cleanly and (ii) pass the benchmark tests.
Table~\ref{tab:rq1_main} summarizes the results.

\noindent\faHandORight~\noindent \textbf{Main results.}
AutoCodeRover resolves 28.79\% (152/528) of the instances while achieving a very high \texttt{Patch Apply Rate} of 96.21\% (508/528).
This suggests that AutoCodeRover is generally reliable at producing syntactically valid patches, but it often fails to fully satisfy the functional requirements enforced by the test suite.
TraeAgent attains a notably higher \texttt{Resolved Rate} of 52.65\% (278/528), indicating stronger task-solving effectiveness.
However, its \texttt{Patch Apply Rate} is substantially lower at 78.98\% (417/528).
mini-SWE-Agent achieves the highest \texttt{Resolved Rate} of 70.08\% (370/528), substantially outperforming both AutoCodeRover and TraeAgent.
At the same time, it maintains a high \texttt{Patch Apply Rate} of 95.83\% (506/528).
This indicates that mini-SWE-Agent combines strong problem-solving capability with stable patch generation, making it the most effective agent on the \appname.

\begin{table}[t]
\centering
\small
\caption{Patch-level results of code agents on the full \appname (528 instances).}
\vspace{-1em}
\begin{tabular}{lcc}
\toprule
\textbf{Agent} & \textbf{Patch Apply Rate} & \textbf{Resolved Rate} \\
\midrule
AutoCodeRover & 508/528 (96.21\%) & 152/528 (28.79\%) \\
TraeAgent & 417/528 (78.98\%) & 278/528 (52.65\%) \\
mini-SWE-Agent & 506/528 (95.83\%) & 370/528 (70.08\%) \\
\bottomrule
\end{tabular}
\label{tab:rq1_main}
\vspace{-1em}
\end{table}

\noindent\faHandORight~\noindent \textbf{Why does TraeAgent have a lower \texttt{Patch Apply Rate}?}
Through manual inspection of failed cases, we find that many apply-check failures are caused by unstable agent-environment interactions rather than incorrect patch intent.
In particular, TraeAgent can enter unrecoverable terminal states during tool use, such as invoking interactive \texttt{Python} sessions or opening files with blocking commands that require explicit keystrokes to exit.
These behaviors may cause the run to hang and prevent clean patch application.

\vspace{-0.5em}
\begin{center}
    \resizebox{\linewidth}{!}{
\begin{tabular}{l!{\vrule width 1pt}p{0.9\columnwidth}}
    \makecell{{\LARGE \faLightbulbO}}  &\textbf{Answer to RQ1:}
On \appname, mini-SWE-Agent has the highest \texttt{Resolved Rate} (70.08\%), outperforming TraeAgent (52.65\%) and AutoCodeRover (28.79\%). TraeAgent’s performance is limited by a lower \texttt{Patch Apply Rate} (78.98\%), highlighting the importance of stable agent–tool interaction.\\
\end{tabular}}
\end{center}

\subsection{RQ2: Reasoning}
\label{rq:rq2}

\begin{table*}[t]
\centering
\small
\caption{Reasoning evaluation results on \appname (mean scores). ``Success''/``Failure'' are split by patch success. We additionally report ``Failure (excl. apply-fail)'' whose patches apply successfully but fail at least one test.
}
\vspace{-1em}
\resizebox{\textwidth}{!}{%
\begin{tabular}{lcccccccc}
\toprule
\textbf{Agent} & \textbf{Group} & \textbf{Score@Goal} & \textbf{Recall@Files} & \textbf{OverPred@Files} & \textbf{Recall@Tasks} & \textbf{OverPred@Tasks} & \textbf{Recall@Steps} & \textbf{OverPred@Steps} \\
\midrule
\multirow{4}{*}{AutoCodeRover} 
& All & 9.344 & 0.840 & 0.067 & 0.675 & 0.255 & 0.397 & 0.579 \\
& Success & 9.397 & 0.961 & 0.034 & 0.853 & 0.136 & 0.565 & 0.447 \\
& Failure & 9.321 & 0.791 & 0.080 & 0.602 & 0.304 & 0.329 & 0.633 \\
& \footnotesize{Failure (excl. apply-fail)} & 9.348 & 0.814 & 0.084 & 0.613 & 0.311 & 0.329 & 0.655 \\
\midrule
\multirow{4}{*}{TraeAgent}
& All & 9.506 & 0.746 & 0.420 & 0.607 & 0.334 & 0.340 & 0.652 \\
& Success & 9.538 & 0.955 & 0.321 & 0.858 & 0.149 & 0.515 & 0.538 \\
& Failure & 9.463 & 0.515 & 0.530 & 0.329 & 0.540 & 0.144 & 0.778 \\
& \footnotesize{Failure (excl. apply-fail)} & 9.574 & 0.640 & 0.568 & 0.472 & 0.377 & 0.214 & 0.742 \\
\midrule
\multirow{4}{*}{mini-SWE-Agent}
& All & 9.251 & 0.890 & 0.049 & 0.751 & 0.208 & 0.445 & 0.584 \\
& Success & 9.261 & 0.954 & 0.039 & 0.841 & 0.141 & 0.527 & 0.508 \\
& Failure & 9.220 & 0.736 & 0.073 & 0.537 & 0.366 & 0.251 & 0.762 \\
& \footnotesize{Failure (excl. apply-fail)} & 9.264 & 0.747 & 0.069 & 0.580 & 0.326 & 0.281 & 0.753 \\
\bottomrule
\end{tabular}%
}
\label{tab:rq2_reasoning}
\vspace{-1.3em}
\end{table*}

While RQ1 focuses on end-to-end patch correctness, RQ2 aims to evaluate the intermediate reasoning processes of code agents when solving repository-level feature addition tasks.
Following the \emph{Reasoning Assessment} workflow described in Section~\ref{sec:evaluationWorkflow}, we first employ an independent summarizer LLM to normalize the intermediate reasoning process (i.e., \emph{Concept Explanations}, \emph{Goal Expectation}, \emph{File Localization}, \emph{Specific Tasks}, and \emph{Task Steps}) of each instance.
If an apply-success patch is generated, we further refine the reasoning chain by applying the same back-inference procedure used to construct the \dx{\emph{Reference Reasoning}} (Section~\ref{sec:reasoninggroundtruthconstruction}), deriving intermediate reasoning directly from the generated patch.
We then compare each agent's reasoning chain against the benchmark-provided \dx{\emph{Reference Reasoning}} using the evaluation metrics introduced in Section~\ref{sec:evaluationMetrics}.
Since modules such as \emph{File Localization} and \emph{Specific Tasks} can be reconstructed and supplemented from the generated patch, apply-fail instances prevent structural reconstruction. To ensure fair comparison, we additionally report the failure subset that excludes apply-fail instances.
Table~\ref{tab:rq2_reasoning} reports the mean reasoning scores for four groups: \emph{All}, \emph{Success} (apply-success and test-success), \emph{Failure} (apply-fail or test-fail), and \emph{Failure (excluding apply-fail)}.
Our key findings are summarized as follows:

\noindent\faLightbulbO~\textbf{Finding 1. Implicit Conceptual Understanding.}
Across the intermediate reasoning process of 528 instances, explicit concept explanations are nearly absent. AutoCodeRover provides them in 0.18\% (1/528) of cases, TraeAgent in 1.70\% (9/528), and mini-SWE-Agent in 0.94\% (5/528). In most instances, agents proceed directly to file exploration or code edits without clarifying domain concepts. This pattern reflects an execution-first reasoning style that is shaped by agent prompting. However, all agents achieve consistently high \texttt{Score@Goal} values (9.2–9.6). To reach such high goal expectation scores, the agents must correctly interpret the domain-specific concepts described in the issue. This indicates that the models understand these concepts internally, even though they rarely explain them explicitly in their intermediate reasoning processes.

\noindent\faLightbulbO~\textbf{Finding 2. Strong Intent Understanding but Waterfall Degradation in Execution.}
Intent comprehension is not the primary difficulty for repository-level code agents. As shown in Table~\ref{tab:rq2_reasoning}, all three agents demonstrate a high-level understanding of feature requests. The \texttt{Score@Goal} metric remains high across agents, with average scores exceeding 9.2 out of 10 (e.g., 9.506 for TraeAgent and 9.344 for AutoCodeRover). Even in failed instances, these goal comprehension scores show a negligible decline. These results indicate that agents can accurately capture high-level intent from natural language issues.
However, the main challenge arises when translating high-level intent into concrete implementation plans. We observe a clear waterfall-style degradation in performance as the reasoning chain progresses from file localization to specific tasks and finally to task steps. For example, although mini-SWE-Agent achieves a strong \texttt{Recall@Files} of 0.890, its coverage decreases to 0.751 for \texttt{Recall@Tasks} and further drops to 0.445 for \texttt{Recall@Steps}. At the same time, \texttt{OverPred@Steps} increases across all models, reaching 0.652 for TraeAgent and 0.584 for mini-SWE-Agent in the overall setting. These findings suggest that while identifying high-level intent is comparatively easy for code agents, accurately planning how to execute it step by step remains a significant challenge.

\noindent\faLightbulbO~\textbf{Finding 3. \dx{Higher Reference Alignment is Associated with Patch Success.}}
By comparing the \emph{Success} and \emph{Failure (excluding apply-fail)} subgroups in Table~\ref{tab:rq2_reasoning}, we observe that successful instances are characterized by high recall and low over-prediction. Specifically, they maintain \texttt{Recall@Files} above 0.950, \texttt{Recall@Tasks} above 0.840 and \texttt{Recall@Steps} above 0.510. In contrast, apply-success but test-fail cases show a decline in recall and an increase in over-prediction. TraeAgent illustrates this pattern with lower \texttt{Recall@Files} (0.955 $\rightarrow$ 0.640) and higher \texttt{OverPred@Files} (0.321 $\rightarrow$ 0.568). Similar trends are observed across other reasoning modules (i.e., tasks and steps).
Averaged across all agents and reasoning modules, apply-success but test-fail instances are characterized by a 35.7\% overall decrease in recall and a 94.1\% increase in over-prediction compared to successful instances.
\dx{These results suggest that functional failure is often associated with lower alignment to the developer-accepted reasoning and a broader modification scope.}

\noindent\faLightbulbO~\textbf{Finding 4. Distinct Reasoning Styles Across Agents.}
The three agents exhibit clearly different reasoning behaviors, reflecting trade-offs between exploration and exploitation.
AutoCodeRover represents a conservative style. It records low over-prediction scores across all reasoning modules. For example, its overall \texttt{OverPred@Files} is only 0.067 and \texttt{OverPred@Tasks} is 0.255. This strict restriction on redundant edits directly explains its high \texttt{Patch Apply Rate} of 96.21\% observed in RQ1. The agent only modifies code entities when it is highly confident. However, this conservative exploitation also limits recall, leading to missed \dx{reference} changes.
In contrast, TraeAgent adopts an exploratory strategy. Although it achieves a high \texttt{Resolved Rate} (52.65\%), its over-prediction is substantially higher, especially at the file and step levels (\texttt{OverPred@Files} = 0.420, \texttt{OverPred@Steps} = 0.652). This pattern indicates that Trae Agent tends to explore more files and task steps, but with weaker control over modification boundaries, which may contribute to its lower \texttt{Patch Apply Rate} (78.98\%).
mini-SWE-Agent demonstrates a more effective balance between coverage and precision. It achieves the highest recall across reasoning modules (\texttt{Recall@Files} = 0.890 and \texttt{Recall@Tasks} = 0.751), while keeping over-prediction relatively low (\texttt{OverPred@Files} = 0.049 and \texttt{OverPred@Tasks} = 0.208). This combination of broad coverage and controlled redundancy contributes to the higher \texttt{Resolved Rate} (70.08\%) observed in RQ1.

\noindent\faLightbulbO~\textbf{Finding 5. Implementation Diversity and Equifinality in Successful Patches.}
Even in successful cases where patches pass all tests, recall at the task and step levels does not reach 1.0. For example, within the \emph{Success} subgroup, \texttt{Recall@Tasks} ranges from 0.841 to 0.858, and \texttt{Recall@Steps} ranges from 0.515 to 0.565. These results indicate that agents can achieve functional correctness without strictly following the \dx{\emph{Reference Reasoning}}.
This reflects a fundamental characteristic of software development: a functional objective can be implemented through multiple implementation paths. An agent can deviate from the \dx{reference} but still produce a successful patch.
\dx{Consequently, the recall metrics in \appname evaluate alignment with developer-accepted reference reasoning rather than strict matching to a single prescribed solution.}

\vspace{-0.5em}
\begin{center}
    \resizebox{\linewidth}{!}{
\begin{tabular}{l!{\vrule width 1pt}p{0.9\columnwidth}}
    \makecell{{\LARGE \faLightbulbO}}  &\textbf{Answer to RQ2:}
Code agents demonstrate strong high-level intent comprehension (average \texttt{Score@Goal} = 9.37/10), but exhibit a waterfall-style degradation when translating goals into specific tasks and plans.
Furthermore, apply-success but test-fail cases are characterized by lower recall (35.7\% decrease) and higher over-prediction (94.1\% increase) compared to successful cases. Finally, agents exhibit distinct exploration–exploitation profiles in their reasoning styles.\\
\end{tabular}}
\end{center}

\begin{table*}[t]
\centering
\small
\caption{Patch-level and reasoning-level results on \appname Lite under different base LLMs. We report the intermediate reasoning metrics for the Failure (excl. apply-fail) group. Here, R@ and O@ denote Recall and OverPrediction, respectively.}
\vspace{-1em}
\resizebox{\textwidth}{!}{%
\begin{tabular}{llccccccccc}
\toprule
\textbf{Agent} & \textbf{Base} & \textbf{Resolved Rate} & \textbf{Apply Rate} &
\textbf{Score@Goal} & \textbf{R@Files} & \textbf{O@Files} &
\textbf{R@Tasks} & \textbf{O@Tasks} &
\textbf{R@Steps} & \textbf{O@Steps} \\
\midrule
\multirow{3}{*}{AutoCodeRover}
& DeepSeek &0.290&0.960&9.358&0.791& 0.056 &0.615 & 0.319 & 0.280 & 0.694 \\
& GLM &0.190&0.960&9.000&0.773& 0.055& 0.585 & 0.257 & 0.295 & 0.681 \\
& GPT&0.370&0.990&9.113&0.769& 0.066& 0.655 &0.238 & 0.362& 0.661 \\
\midrule
\multirow{3}{*}{TraeAgent}
& DeepSeek &0.520&0.800&9.731&0.665& 0.415& 0.558& 0.361 & 0.284 & 0.735 \\
& GLM &0.720&0.920&8.850&0.677& 0.378& 0.549& 0.347& 0.273& 0.729 \\
& GPT&0.770&0.900&9.333&0.767& 0.000& 0.688& 0.156& 0.221& 0.672\\
\midrule
\multirow{3}{*}{mini-SWE-Agent}
& DeepSeek &0.710&0.970&9.238&0.760& 0.019 & 0.637 & 0.197& 0.333 &0.667 \\
& GLM &0.790&0.990&8.222&0.664& 0.127& 0.521 & 0.215 & 0.333 & 0.600 \\
& GPT&0.710&0.990&9.890&0.749& 0.033 & 0.745 & 0.213 & 0.400 & 0.590\\
\bottomrule
\end{tabular}%
}
\vspace{-1em}
\label{tab:rq3_reasoning}
\end{table*}

\subsection{RQ3: Generality}
\label{rq:rq3}
To understand how the underlying LLM influences agent behavior, we evaluate three agents under different base LLMs. 
Beyond reporting overall performance, we analyze the intermediate reasoning process of \emph{Failure (excluding apply-fail)} cases to investigate why patches fail the tests.
The results are shown in Table~\ref{tab:rq3_reasoning}.

\noindent\faThumbsUp ~\textbf{Finding 1. Base LLM Impact Depends on Agent Architecture.}
The choice of the base LLM leads to performance variations. For example, the \texttt{Resolved Rate} of AutoCodeRover varies from 0.190 (\texttt{GLM}) to 0.370 (\texttt{GPT}), while TraeAgent shows a wider variation from 0.520 (\texttt{DeepSeek}) to 0.770 (\texttt{GPT}). In contrast, mini-SWE-Agent remains relatively stable across models, with \texttt{Resolved Rate} ranging from 0.710 to 0.790.
These results indicate that no single base LLM consistently performs best across agents, and the agent framework determines the operational boundary of the base LLM. 
In conservative agents such as AutoCodeRover, strict AST-guided navigation restricts exploration and causes different LLMs to exhibit similar reasoning behaviors. For instance, AutoCodeRover shows consistent reasoning patterns across models, with \texttt{Recall@Steps} ranging only from 0.280 to 0.362 and \texttt{OverPred@Steps} remaining high (0.661–0.694). This indicates that the reasoning trajectory is largely determined by the agent framework rather than the underlying model.
In contrast, a more balanced architecture can reduce the system’s sensitivity to the base LLM. 
mini-SWE-Agent shows relatively stable performance across models, achieving \texttt{Resolved Rate} of 0.710 (\texttt{DeepSeek}), 0.790 (\texttt{GLM}), and 0.710 (\texttt{GPT}) while maintaining high \texttt{Patch Apply Rate} (0.970–0.990). 
This finding suggests that well-designed agent architectures can mitigate differences in model capability and achieve more robust cross-model performance.

\noindent\faThumbsUp ~\textbf{Finding 2. Stronger Base LLMs Improve Different Agents Through Different Mechanisms.}
The benefits of stronger base LLMs vary depending on the agent design. For conservative agents such as AutoCodeRover, improvements from \texttt{GPT} primarily appear on the recall side, increasing coverage of relevant tasks and steps. For example, \texttt{Recall@Steps} increases from 0.280 to 0.362, while the reduction in over-prediction remains relatively small (\texttt{OverPred@Steps} decreases from 0.694 to 0.661). This pattern indicates that stronger models help AutoCodeRover identify more required modifications.
In contrast, exploratory agents benefit through improved redundancy control. For TraeAgent, switching to \texttt{GPT} reduces unnecessary exploration. \texttt{OverPred@Files} drops from 0.415 (\texttt{DeepSeek}) to 0.000, while \texttt{OverPred@Tasks} decreases from 0.361 to 0.156. These results suggest that stronger base LLMs improve exploratory agents mainly by reducing redundant modifications.

\noindent\faThumbsUp ~\textbf{Finding 3. Base LLMs Exhibit Distinct Reasoning Styles.}
Different base LLMs consistently exhibit distinct behavioral patterns. \texttt{GPT} tends to produce more precise modifications, resulting in lower over-prediction across multiple settings. This effect is most evident in TraeAgent, where \texttt{OverPred@Files} decreases from 0.415 to 0.000 with \texttt{GPT}, while \texttt{OverPred@Tasks} declines from 0.361 to 0.156. These results suggest that \texttt{GPT} is more effective at restricting the modification scope and avoiding unnecessary exploration.
In contrast, \texttt{GLM} consistently achieves the lowest \texttt{Score@Goal} across all agents, averaging 7.9\% lower than the other base LLMs. This pattern indicates weaker alignment with the overall task objective, making \texttt{GLM} more prone to deviating from the intended feature goal. Overall, these results suggest that different base LLMs introduce distinct reasoning styles into repository-level code agents.

\vspace{-0.5em}
\begin{center}
\resizebox{\linewidth}{!}{
\begin{tabular}{l!{\vrule width 1pt}p{0.9\columnwidth}}
\makecell{{\LARGE \faLightbulbO}} & \textbf{Answer to RQ3:}
The choice of the base LLM affects agent performance on \appname, but its impact is influenced by the agent architecture. For well-designed agents such as mini-SWE-Agent, performance remains stable across base LLMs. Moreover, different base LLMs exhibit distinct reasoning styles: \texttt{GPT} tends to produce more precise modifications, whereas \texttt{GLM} shows weaker goal alignment.\\
\end{tabular}}
\end{center}

\section{Discussion}
\label{sec:discussion}
\subsection{Limits of Tests and Reference Reasoning}
\label{sec:LimitsofTests}
\dx{The reference reasoning in \appname is derived from developer-accepted patches. However, feature requests define expected behavior rather than a unique implementation. A test-passing patch may use a different implementation path while still satisfying the request. We examine this limitation from three aspects: whether the exposed FTP tests cover the feature request, whether the reference reasoning can be derived from agent-visible inputs, and whether successful patches use alternative implementation paths.}

\noindent\dx{\textbf{Executable Specification Coverage.}
Prior work shows that developer tests may not fully capture issue semantics~\cite{wang2026solvedissues}. Because \appname exposes \emph{FTP} tests to agents, we examine how well these tests cover the corresponding \emph{Feature Requests}. We use \texttt{GPT-5.5} to compare each issue description with its exposed \emph{FTP} tests and classify the coverage as full, partial, or insufficient. Among the 528 instances, 499 (94.5\%) have full or partial coverage, while 29 (5.5\%) are insufficient. \appname includes \emph{FTP} tests in the prompt as a shared specification of expected behavior. Our goal is therefore to evaluate whether agents can solve feature addition tasks with developer-provided tests. Although these tests may not capture every functional requirement, they provide a practical basis for evaluating agents under this setting.}

\noindent\dx{\textbf{Visible Input Derivability.}
Because parts of the \emph{Reference Reasoning} are inferred from the \emph{Gold Patch}, we assess how much of it can be derived from the agent-visible inputs. Since \appname targets feature-addition tasks where agents receive both the \emph{Feature Request} and \emph{FTP} tests, we perform this analysis under the same setting using 100 sampled instances. Two reviewers independently inspect the \emph{Feature Request} and \emph{FTP} tests to determine whether each reference \emph{Specific Task} and \emph{Task Step} can be derived from these inputs.
They also identify required tasks or steps that are supported by the inputs but absent from the \emph{Reference Reasoning}.
Disagreements are resolved through discussion.
Overall, 174 of 185 \emph{Specific Tasks} (94.05\%) and 251 of 256 \emph{Task Steps} (98.05\%) are supported by the visible inputs.
The reviewers also identify six required \emph{Specific Tasks} and two \emph{Task Steps} that are absent from the \emph{Reference Reasoning}, affecting eight instances.
These results show that the \emph{Reference Reasoning} is largely grounded in agent-visible inputs and support its use as a diagnostic signal for reasoning alignment.}

\noindent\dx{\textbf{Implementation Diversity.}
To examine whether successful patches always follow the developer-accepted reference trajectory, we analyze 100 sampled instances with at least one successful patch, yielding 457 successful records. Among them, 422 (92.3\%) fully cover the reference \emph{Relevant Files}, and 342 (74.8\%) fully cover the reference \emph{Specific Tasks}. The remaining 127 records (27.8\%) pass patch evaluation but do not fully align with the reference trajectory. We use GPT-5.5 to classify these cases and find that 89 (70.1\%) reflect meaningful implementation diversity. The most common pattern is \emph{alternative minimal path}, where the agent satisfies the feature request through a smaller or different implementation than the reference. For example, in a pytest instance, the agent reuses the existing mechanism to pass timing fields into \texttt{TestReport}, while the reference modifies \texttt{TestReport.\_\_init\_\_} directly. These results show that imperfect alignment does not necessarily indicate incorrect reasoning. The reasoning metrics in \appname measure alignment with a developer-accepted reference trajectory rather than absolute correctness.}

\subsection{Design Insights for Code Agents}
\dx{Beyond comparing agent performance, \appname provides diagnostic signals that reveal reasoning failures in agents. These signals offer practical design principles and guide improvements for agents.}

\noindent\dx{\textbf{Design Insights.}
\appname provides practical design principles for repository-level code agents.
\emph{(1) Avoid overly narrow search.}
AutoCodeRover uses AST and API information to guide repository navigation.
It achieves a high \texttt{Patch Apply Rate} of 96.21\% and low over-prediction, but its \texttt{Resolved Rate} is only 28.79\% and its recall is low.
This result suggests that conservative exploration can produce stable patches,
but may miss files or changes required to complete the task.
\emph{(2) Control redundant modification.}
TraeAgent uses iterative exploration and test-time scaling.
It achieves a higher \texttt{Resolved Rate} of 52.65\%, but has a much
higher \texttt{OverPred@Files} of 0.420.
This result suggests that broad exploration should be combined with stronger
control over the modification scope.
\emph{(3) Maintain robust tool states.}
Manual inspection shows that some TraeAgent failures arise from unstable
tool interactions, such as interactive sessions and blocking commands.
In contrast, mini-SWE-Agent uses a lightweight bash-centered interaction loop that keeps the interaction workflow simple and stable. These suggest that repository-level agents should manage tool states to keep the reasoning and action loop executable.}

\noindent\dx{\textbf{Benchmark-guided Case Study.}
To examine whether the reasoning metrics can guide practical improvements, we conduct a case study on TraeAgent with \texttt{DeepSeek} as the base model. RQ2 shows that apply-success but test-fail cases often have low file-level alignment (\texttt{Recall@Files}=0.640).
Manual inspection reveals a consistent pattern. TraeAgent often starts editing after finding the first plausible implementation file and stops exploring the repository too early. When the initial edit introduces new errors, it follows stack traces for local debugging instead of returning to file localization, leaving other required files undiscovered.
Based on this observation, we add a lightweight \emph{Repository-Level File Localization} stage before code editing. The agent first summarizes signals from the issue description and FTP tests, identifies candidate implementation files, and records the final localization decision. The remaining workflow is unchanged.
On 16 representative localization-failure cases, the modified agent improves \texttt{Recall@Files} from 0.496 to 0.565, increases perfect file localization from 0\% to 18.8\%, and resolves four previously failed instances. We also evaluate 10 previously successful cases and observe regressions in some of them, mainly due to TraeAgent becoming stuck during long-text interactions. Our goal is not to build a stronger agent, but to validate whether \appname can identify actionable reasoning bottlenecks. The improvements on previously failed cases show that the file localization is a key bottleneck and the corresponding diagnostic signal can guide architectural improvements.}

\subsection{Threats to Validity}
\subsubsection{Internal Validity}
A potential threat arises from the use of the \emph{LLM-as-Judge} in reasoning assessment, which may introduce evaluator bias in semantic matching. To mitigate this risk, we adopt a hybrid evaluation strategy. For tasks involving structured outputs (e.g., \emph{Specific Tasks}), we combine deterministic matching rules with semantic similarity checks, thereby reducing reliance on purely LLM-based judgments.
Another threat comes from the construction of the \dx{\emph{Reference Reasoning}}, as several reasoning modules are generated with LLM assistance and may therefore contain inaccuracies or omissions. To mitigate this threat, we complement LLM-assisted construction with patch-driven analysis and further perform an LLM-assisted audit over all instances, followed by targeted human validation (Section~\ref{sec:human_validation}) to verify the semantic correctness and consistency of the resulting \dx{\emph{Reference Reasoning}}.
Another threat concerns the LLM-based classifications used in Section~\ref{sec:LimitsofTests}. Although these analyses provide supporting evidence, their classifications have not been independently validated by human reviewers. Future work may further assess their reliability through manual checks.
A further threat concerns potential randomness in agent evaluation. In our experiments, we report results from a single run for each instance, which may introduce variance. However, this design follows the \emph{first-attempt} evaluation setting commonly adopted in existing agent benchmarks~\cite{li2025fea,jimenez2023swe}, reflecting realistic deployment scenarios in which agents are expected to solve tasks without repeated retries.
In addition, we set the generation temperature to 0 to ensure more stable and reproducible results.

\subsubsection{External Validity}
A threat to external validity is that our experiments cover only a limited set of agents and base models. To mitigate this threat, we select agents with diverse design philosophies and base models from different providers.
Future work may extend the evaluation to additional agents and models as the ecosystem continues to evolve.
Another threat is that our benchmark evaluates only the explicit intermediate reasoning exposed by agents, whereas agents may also rely on implicit reasoning that is not directly observable. To mitigate this limitation, we refine intermediate reasoning by analyzing the final code patches generated by the agents. This approach allows us to approximate the underlying reasoning process even when parts of it are not explicitly reported.
The final threat arises from differences in how agents expose their intermediate reasoning processes, which may affect the fairness of cross-agent comparison. To mitigate this threat, we map the intermediate reasoning process into unified reasoning modules, making the evaluation more comparable across different agent frameworks.

\section{Related Work}
\label{sec:relatedwork}
\subsection{Code Generation Benchmarks}
Code generation is a core task in software engineering, and a wide range of benchmarks have been proposed to evaluate the code generation capabilities of LLMs.
Early benchmarks focus on single-function synthesis from natural language descriptions and unit tests, such as HumanEval~\cite{chen2021evaluating} and MBPP~\cite{austin2021program}. Subsequent benchmarks extend this paradigm by introducing more complex problems and richer test suites, including APPS~\cite{hendrycks2021measuring}, EvalPlus~\cite{liu2023your}, and CoderEval~\cite{yu2024codereval}. 
Other benchmarks, including ClassEval~\cite{du2023classeval}, BigCodeBench~\cite{zhuo2024bigcodebench}, and FullStackBench~\cite{cheng2024fullstack}, further expand the evaluation scope to class-level abstractions, large-scale codebases, and full-stack development scenarios.
Despite these advances, most benchmarks still ignore repository context and evaluate code generation in isolation, limiting their ability to reflect realistic scenarios where models must reason over large repositories and generate multi-file changes.
To address this gap, recent work has proposed repository-level code generation benchmarks, such as FEATBench~\cite{chen2025featbench}, DevEval~\cite{li2024deveval}, EvoCodeBench~\cite{li2024evocodebench}, CodevBench~\cite{pan2024codev}, and ExecRepoBench~\cite{yang2024execrepobench}.
Another line of work constructs benchmarks directly from real-world issue–PR pairs.
SWE-bench~\cite{jimenez2023swe} extracts bug-fixing tasks from GitHub issues and corresponding patches, enabling end-to-end evaluation of automated software maintenance.
Similarly, FEA-bench~\cite{li2025fea} focuses on feature addition tasks derived from real repositories, capturing the challenges of introducing new functionality while preserving existing behavior.
However, these benchmarks primarily focus on whether the final generated patch passes the test suite, while largely ignoring the quality of the intermediate reasoning process.
In contrast, our benchmark augments repository-level tasks with structured \dx{\emph{Reference Reasoning}}, enabling systematic evaluation of intermediate reasoning \dx{alignment} and facilitating fine-grained error diagnosis.

\subsection{Intermediate Reasoning Evaluation}
\dx{Recent work has increasingly recognized that evaluating intermediate reasoning provides diagnostic signals beyond final-answer correctness. Existing work mainly focuses on assessing the quality of reasoning traces in natural language and mathematical reasoning. For example, ROSCOE~\cite{golovneva2022roscoe} measures semantic consistency, logicality, and informativeness of step-by-step rationales, ReCEval~\cite{prasad2023receval} evaluates whether each reasoning step is supported by the input and preceding context, and ProcessBench~\cite{zheng2025processbench} models on identifying the earliest erroneous reasoning step in mathematical solutions.
Reasoning evaluation has also been extended to code and agent settings. CRUXEval~\cite{gu2024cruxeval} evaluates reasoning over short executable functions, while Xue et al.~\cite{xue2026empirical} analyze the efficiency, logical consistency, and completeness of reasoning traces for BigCodeBench tasks. AgentBoard~\cite{ma2024agentboard} introduces fine-grained evaluation for multi-turn agents in general interactive environments. More closely related to repository-level software engineering, EnConda-Bench~\cite{kuang2025process} provides process-level diagnosis for environment configuration, and AgentLens-Bench~\cite{sahoo2026agentlens} constructs process references from multiple successful repair trajectories to evaluate trajectory coherence, efficiency, and divergence.
However, existing work either evaluates intermediate reasoning outside repository-level software engineering or focuses on specialized software engineering subtasks rather than feature implementation. Repository-level benchmarks such as SWE-bench and FEA-bench evaluate only the final patch correctness without structured reasoning references. In contrast, \appname combines executable patch verification with structured developer-accepted \emph{Reference Reasoning}, enabling joint evaluation of final patch correctness and intermediate reasoning alignment for repository-level feature addition.}

\section{Conclusion}
\label{sec:conclusion}
In this paper, we present \appname, a reasoning-augmented benchmark for evaluating repository-level code agents on real-world feature addition tasks. \appname contains 528 instances from 12 open-source GitHub repositories, each paired with executable verification and structured \dx{\emph{Reference Reasoning}}, enabling evaluation at both the patch level and the intermediate reasoning level.
Our experiments on three representative agents show substantial differences in end-to-end capability, with \texttt{Resolved Rate} ranging from 29\% to 70\%. More importantly, the results demonstrate that evaluating only final test outcomes is insufficient for understanding agent behavior. Reasoning-level analysis reveals a clear waterfall-style degradation: agents generally perform well at high-level goal understanding, but become much weaker when translating goals into files, tasks, and especially concrete implementation steps. Step-level reasoning is the most critical bottleneck, and test-fail instances are often associated with lower recall (a 35.7\% decrease) and excessive over-prediction (a 94.1\% increase).
We further find that the impact of the base LLM is not uniform across agents. Instead, it is strongly shaped by the agent architecture, suggesting that model capability and agent design should be studied jointly rather than in isolation. 
Overall, \appname provides a practical foundation for diagnosing, comparing, and improving repository-level code agents beyond black-box patch success evaluation.

\vspace{-0.15cm}
\section{Data Availability Statement}
\label{sec:dataAvailability}
The dataset and the code are publicly available on our website~\cite{zenodo_race2026}.

\vspace{-0.15cm}
\begin{acks} 
This research is supported by National Key R\&D Program of China (No. 2024YFB4506400) and the CCF-Huawei Populus Grove Fund.
\end{acks} 

\balance
\bibliographystyle{ACM-Reference-Format}
\bibliography{references} 

\end{document}